**Biaxial layering transition of hard rod-like particles in narrow slit-like pores**

Hamdollah Salehi [a], Sakine Mizani [a], Roohollah Aliabadi [b,*] and Szabolcs Varga [c]

[a] *Department of Physics, Faculty of Science, Shahid Chamran University of Ahvaz, Ahvaz, Iran*

[b] *Department of Physics, Faculty of Science, Fasa University, 74617-81189 Fasa, Iran*

[c] *Institute of Physics and Mechatronics, University of Pannonia, P.O. Box 158, Veszprém, H-8201 Hungary*

**Abstract**

The phase behaviour of hard rectangular rods with length *L* and diameter *D* is studied in a narrow slit-like pore using the Parsons-Lee density functional theory. Using the restricted orientation approximation, we find strong adsorption at the walls with planar ordering, second order uniaxial-biaxial ordering transitions and first order layering transitions. The layering transition takes place between two fluids having *n* and *n*+1 layers, where the layer spacing is in the order of *D*. In the case of weak shape anisotropy (*L/D*=3), the coexisting fluids can be either uniaxial or biaxial, while both phases are found to be biaxial for *L/D*=6 and *L/D*=9. Interestingly, even two or more layering transitions can be observed with increasing density at a given shape anisotropy and pore width.

**Introduction**

Over the years, anisotropic colloidal particles have been investigated experimentally [1, 2, 3, 4], theoretically [5, 6, 7, 8] and using computer simulations [9, 10]. Despite numerous studies on anisotropic colloidal particles, the bulk and confined properties of colloidal systems are still very interesting from both scientific and industrial point of views due to their stimulating phase behaviors and to the convenient alignment of the anisotropic molecules by surfaces and external electric (magnetic) fields [11, 12]. Lyotropic liquid crystals are one group of colloidal particles consisting of high-molecular-weight particles where the mesophase structures originate from gaining entropy only [13]. They even show much richer and fascinating phase behavior in pores because the confinement gives rise to different anchoring and surface induced ordering phenomena like capillary nematization, smectic A, columnar and T phases [6, 12, 14, 15]. The observed rich phase behavior is purely due to the presence of different particle-particle and particle-wall interactions and to the competition between them [16]. The understanding and predicting the self-assembly of colloidal nanoparticles (including the metallic and semiconducting nanorods) in nanopores are very important theoretical problems due to the widespread application of nanoelectronic devices [17, 18, 19]. In this regard, hard rods such as the hard parallelepipeds with square or different cross sections can serve a good model system to understand the ordering properties of confined metallic nanorods, which have been synthesized using different materials [20].

[*] aliabadi313@gmail.com, aliabadi@fasau.ac.ir

Hard rods placed between two planar hard walls accumulate in planar order at the walls and exhibit a surface phase transition between isotropic and nematic films, where the nematic wetting is complete [6, 21, 22]. The high density smectic and solid structures of hard rods are also affected by the presence of walls. For example, finite and infinite sequence of layering phenomena were reported for a smectic film near an attractive wall [23, 24]. A layering transition, where the particles are in smectic order, was predicted for confined hard spherocylinders in the high density regime by de las Heras et al. [25]. They show that in wide pores there is a first order layering transition from a smectic phase with $n$ layers ($S_n$) to $n+1$ smectic layers ($S_{n+1}$) where the $S_n$, $S_{n+1}$ and nematic phases meet and a triple point is established by these three phases. Whereas this situation is different for enough narrow pores where the triple points vanish and the layering transition terminates at a critical point [25]. First order layering transitions were also observed for confined biaxial hard particles using the Onsager's second virial theory [26]. The presence of layering transition was also investigated in a mixture of hard parallel cylinders with the same diameter but different lengths near a hard wall [27] and between two parallel walls [18]. In addition to the layering transitions a demixing between two phases was observed, where one phase is rich in short, while the other in long rods [18]. The entropic wetting of a binary mixture of hard spheres near a hard wall using fundamental measure theory was also investigated [28]. Depending on the value of the hard wall offset parameter, this system shows wetting and layering transition. Layering behaviors have been also observed in colloidal-polymer mixtures [29, 30, 31] and in the system of confined two dimensional hard rods [32, 33, 34, 35].

In this paper, we study the surface induced biaxial ordering and the layering transitions of hard parallelepipeds in slit-like pores, where $L/D$ is constant and the wall-to-wall separation is changed. We determine the structure and the possible phase transitions at the following three different shape anisotropies: *L/D*=3, *L/D*=6 and *L/D*=9. According to our best knowledge the layering transition has not been observed between two biaxial nematic phases, where the thickness of the layers is in the order of *D*. Here, this type of phase transition is reported for the first time.

**Theory**

The phase behaviors of confined hard rectangular rods with length *L* and side length *D*, which are placed between two parallel hard walls, are investigated using the Parsons-Lee modification of the second virial density functional theory [12, 36]. The hard walls are located at *z*=0 and *z*=*H* positions and they are infinite in the *x-y* plane. We use the three-state restricted orientation approximation, where the particles are allowed to orient along *x*, *y* and *z* directions. In addition to this, we do not consider the high density in-plane crystalline ordering, i.e. the local density of orientation $i$ ($\rho_i$, where $i = x, y, z$) is assumed to depend on *z* coordinate only.

As the particle-particle and the particle-wall interactions are hard repulsion, the system is athermal and the temperature (*T*) does not play a role in the phase behavior of the system. Therefore our confined system is entropy (*S*) driven and the minimum of the free energy (*F*) determines the equilibrium phase behavior (*F~-TS*). We can write the free energy as a sum of ideal, excess and external terms $(F = F_{id} + F_{exc} + F_{ext})$. The ideal free energy term prefers the homogeneous distribution of the particles in both orientation and position and it is given by

$$\frac{\beta F_{id}}{A} = \sum_{i=x,y,z} \int dz \rho_i(z)[\ln \rho_i(z) - 1], \quad (1)$$

where $\beta$ is the inverse temperature and *A* is the surface area. The excess free energy term is responsible for the stabilization of the ordered phases since it has minimum for such a structure where the excluded volume between two particles is minimal. We can write this part as follows

$$\frac{\beta F_{exc}}{A} = \frac{1}{2} c \sum_{i,j=x,y,z} \int dz_1 \rho_i(z_1) \int dz_2 \rho_j(z_2) A_{exc}^{ij}(z_1 - z_2), \quad (2)$$

where $c = (1 - 3\eta/4)(1-\eta)^{-2}$, $\eta$ is the packing fraction and $A_{exc}^{ij}$ is the excluded area between two particles with orientations $i$ and $j$. The last term, which restricts the particles to be located between the two hard walls, can be written as

$$\frac{\beta F_{ext}}{A} = \sum_{i=x,y,z} \int dz \rho_i(z) \beta V_{ext}^i(z), \tag{3}$$

where $V_{ext}^i$ is the external potential, which is infinite if a particle with orientation $i$ overlaps with the walls and it is zero otherwise. We determine the equilibrium local density at a given packing fraction ($\eta$) and the wall-to-wall distance ($H$). Here we do not present the minimization procedure of the free energy and the applied numerical methods, because these have been already written down in our previous publication [12]. We take $D$ as a unit of the length, i.e. $z^* = z/D$ is the dimensionless distance and $\rho^* = \rho D^3$ is the dimensionless density. The resulting local densities of different structures obtained from the minimization of the free energy and the phase diagrams are presented in the following section.

**Results**

We examine the orientational and positional ordering of hard rectangular rods in such a narrow slit-like pore, where the wall-to-wall distance between the two planar hard walls allows the formation of only one layer in homeotropic order (the long axes of the particles are along the $z$ axis), which be achieved with the condition of $L<H<L+D$. However, several layers can be realised in planar order as the rods are intermediately long ($3D<L<9D$). The lowest limit of the number of planar layers is two, because the planar adsorption at the two walls is always present, while the upper limit is the integer of $H/D$, which does not exceed nine even for the widest pore ($H_{max} = 10D$). Therefore, all of the examined systems can be considered as quasi-two-dimensional (q2D), because uniform fluid structures cannot develop in the middle of the pore at intermediate and high densities. For example, in order to detect capillary nematisation the wall-to-wall distance ($H$) should be about three times longer than the length of the rod ($L$). The competing structures with such conditions are the planar isotropic, the nematic and the layered. The planar isotropic is a q2D isotropic phase, where the adsorbed particles are parallel with the walls, but there is no in-plane orientational order. The nematic one is actually a biaxial or q2D nematic phase, because the planar order is accompanied by in-plane orientation order. Finally the layered structure can be either homoetropic monolayer, where the rods are oriented perpendicularly to the walls, or it consist of $n$ layers of planar monolayers. As we choose $L/D=3$, 6 and 9 in this study, the close packing structure is degenerated. This is due to the fact that the homeotropic and planar layers can fill the 2D plane perfectly and the out-of-plane space filling is the same for both layered structures. This means that the maximum of the packing fraction is the same for both structures, i.e. $\eta_{max} = L/H = \text{int}(H/D)/H$ for all integer $L/D$ and $L<H<L+D$. Even though the homeotropic monolayer structure can be very competitive with the planar ones in very narrow pores ($H<3D$) [12], the translational entropy is now more dominant and works against the homeotropic ordering. This is simple the consequence of the available distance along the pore, which is $H-D$ in the planar, while it is only $H-L$ in the homeotropic ordering. As a result the translational entropy shifts the homeotropic order into the region of the packing fraction, which can be above its close packing value ($\eta_{max}$). Note that our approximate DFT does not take into account that $\eta$ must be lower than $\eta_{max}$ as the free energy functional diverges at $\eta =1$ and not at $\eta = \eta_{max}$ (see $c$ in Eq. (2)).

Now we continue with the phase diagram of the confined hard rods with $L/D=3$ and pore width $3<H/D<4$, which is presented in Fig. (1). As we do not examine the stability of possible solid phases, we show the phase diagram below $\eta =0.7$. This value is very close to $\eta_{max}(H/D=4)=3/4$, but a bit far from $\eta_{max}(H/D=3)=1$. At low densities ($\eta$) the phase is isotropic with planar ordering at the walls and there are homeotropically ordered particles in the middle of the pore, which is due to the gain in the entropy of mixing and packing entropy. As the planar ordering is very efficient in maximization of the available volume (or minimization of the excluded volume) at the walls, the wall induced adsorption creates such a high density at the walls, that the excluded area can be lowered substantially with the in-plane nematic ordering at intermediate packing fractions. Therefore a second order q2D isotropic-nematic transition occurs, which corresponds to planar isotropic (I)-biaxial nematic (BN) transition in three dimensions. The structure of planar isotropic ($\rho_x(z)=\rho_y(z)$) and the biaxial nematic ($\rho_x(z) \neq \rho_y(z) \neq \rho_z(z)$) phases are shown in Figs. 1(b) and 1(c). Fig. 1(a) shows that the I-BN transition density decreases with increasing $H$ if the number of planar layers is two. The reason for this is that the fraction of particles staying in homeotropic order decreases with widening pore. This is crucial because the particles staying in homeotropic order are nonmesogenic as they have square cross section. In wider pores planar ordering with three layers emerges, where the peaks are narrower and the particles of the opposite walls do not overlap with each other [see Figs. 1(d) and 1(e)]. This way of packing makes possible to stabilize the planar isotropic order at higher densities. Therefore the orientational ordering transition between three-layer isotropic (3LI) and three-layer biaxial nematic (3LBN) is shifted to higher packing fractions and it is second order. The I-BN and 3LI-3LBN transitions are not connected, but a first order phase transition separates them from each other. Depending on the value of $H$ this first order transition takes place between different structures: BN-3LBN for $3<H/D<3.28$, BN-3LI for $3.28<H/D<3.34$ and I-3LI for $3.34<H/D<3.51$. Note that re-entrant phenomenon occurs for $3.28<H/D<3.34$ as I-BN-3LI-3LBN phase sequence take place with increasing density. Here both the isotropic and the nematic phases are re-entrant. The low density isotropic phase with two planar layers and the high density isotropic phase with three planar layers is interrupted with biaxial nematic order. In the case of biaxial order there is a 3LI phase between the low and high density biaxial nematic phases. The I-3LI transition terminates in a critical point at $H/D=3.51$. At wider pores the isotopic phase can transform continuously into 3LI as there are less particles with ordering along $z$ axis. To prove that the order of phase transition is related with the amount of homeotropically ordered rods, we show the results of $\rho_x(z=H/2)=\rho_z(z=H/2)$ equation as a dashed curve in Fig. 1 (a), which turns out to be the continuation of phase transition. In wider pores the fraction of particles with homeotropic orientation decreases substantially, because the homeotropic particles interacts with all planar layers, which has high excluded volume cost. Therefore the planar ordering becomes stronger and more planar layers evolve in wide pores. This can be seen in Fig. 2, where $L/D=6$ and $6<H/D<7$. One can see that as the particles are more anisotropic, the planar isotropic-biaxial nematic transition occurs at lower packing fractions, which depends very weakly on $H$ (see Fig. 2(a)). Now the number of planar layers can be four or five at the orientational ordering transition even if six layers can accommodate into the pore. The typical biaxial phases with four and five planar layers are shown in Figs. 2(b) and 2(c). With increasing density the appearance of a new planar layer can decrease further the free energy such a way that the excluded volume gain is more than the loss in the transitional entropy. The emergence of a new planar layer is usually not continuous, because it requires the movement of the existing layers to the direction of the walls, which produces less room for the existing layers. If this structural change is accompanied by increasing the free energy at some densities, a first order phase transition connects the old and the new structures. This happens here, where a layering transition takes place between two biaxial nematic phases, where the number of planar layers changes from four to five or five to six. It can be seen in Fig. 2(a) that the pore is enough wide for continuous structural change from four to five planar layers if $H/D>6.27$. The emergence of the new peak in the middle of the pore can be examined with the change of the density profile in the vicinity of the centre of the pore as follows $\rho_x(z=H/2)=\rho_x(z=H/2-\Delta z)$, where $\Delta z$ is the grid size in the numerical calculation. This equation provides the packing fraction of the structural change, which is shown with dotted curve in Fig. 2(a). It is interesting that even two layering transition emerge with increasing packing fraction for $H/D<6.27$. In addition to this, both 5LBN and 6LBN phases can be destabilized with narrowing the pore because the accommodation is harder with more layers. The phase diagram of $L/D=9$ is very similar, but it more complex due to the increasing number of layered phases. The wall induced biaxial nematic order is stabilized at very low packing fraction and three layering transitions take place between biaxial nematic phases having $n$ and $n+1$ planar layers ($n=6,7$ and 8). Fig. 3(a) shows that the all layering transition weakens with increasing $H$ and the layering with 6 and 7 layers terminates in a critical point. Our results show also that all layering transitions have finite range of $H$ for the existence. In addition to this the range of $H$, where an $n$-$n+1$ layering transition exists, widens with increasing $n$. Therefore even more layered phases can exist for $H/D>10$ even if the minimum number of layers also increases with $H$. Interestingly, the accommodation into seven layers is still hard for $9<H/D<9.13$ even if there are room even for nine layers. However, the change in the number of planar layers can be done continuously with increasing density at wider pores, because the local density is almost constant in neighbourhood of the pore's centre and the new peak can emerge easily in the middle of the pore. This can be seen in Figs. 3(b) and 3(c), where the density profiles indicate six and seven layers at $H/D=9.2$, respectively. The formation of eight layers from seven is harder, because the central layer of the 7LBN structure should split into two peaks. It is even harder to find a room for the 9th layer in the 8LBN structure. As a result both 7-8 and 8-9 layering transitions are first order. From these results we expect that more and more layering transitions emerge with increasing shape anisotropy if the wall-to-wall distance satisfies the $L<H<L+D$ condition.

**Summary**

We have shown that a wall induced biaxial nematic ordering and a layering transition involving planar isotropic and biaxial nematic phases can exist if the particle-particle and the particle-wall interactions are hard repulsive. It is demonstrated that the rod-like particles prefer the planar ordering and they are adsorbed at the walls due to the excluded volume gain by the planar adsorption. At higher densities the biaxial nematic ordering emerges as the surface density of the adsorbed particles exceeds the isotropic-nematic transition density of 2D hard rectangles. The biaxial ordering occurs at lower densities with increasing shape anisotropy due to the higher packing entropy gain over the orientational entropy loss. The mechanism of the layering transition is more complicated because the biaxial nematic phase is inhomogeneous between the two planar walls. As the walls are close to each other, the inhomogeneous fluid structure cannot relax to the bulk value in the middle of the pore and the interference of the wall-effects determine the number of layers. We have observed that two planar fluid layers are always present and they are located the walls, but a homeotropically ordered layer competes with the planar one in the middle of the pore. At weak shape anisotropy ($L/D=3$) we have found that the middle of the pore is dominated by a fluid layer with homeotropic order, which becomes weaker with widening pore width. This is due to the fact the formation of an extra planar layer is easier in wider pores. The homeotropic layering becomes very weaker for $L/D=6$ and negligible for $L/D=9$, because the available room for the particle's centre is still in the order of $D$ in homeotropic order ($H-L$), while it becomes very large in planar one ($H-D$) as $L<H<L+D$. As the number of planar layers cannot exceed the integer of $H/D$, the number of layering transitions is limited. We have detected layering transitions between biaxial nematic phases having $n-1$ and $n$ layers, where $n=3$ for $L/D=3$, $n=5$ and 6 for $L/D=6$ and $n=7$, 8 and 9 for $L/D=9$. It is also shown that $n-1$ to $n$ layering transition terminates at a critical pore width ($H_{cr}$) as the formation of a new peak or a split of the existing peak into two in the middle of the pore can be realised with less packing entropy cost in wider pores.

Our choices for the shape anisotropy and pore width are specific in a sense that mixed planar-homeotropic phases, which consist of $n$ planar layers and $m$ homeotropic ones, are not allowed to form. If we allow $H$ to be greater than $L+D$, the number of possible new structures increases as we have shown in our previous study [12]. It is also true that the present restricted orientation approximation and the neglect of possible solid phases have both qualitative and quantitative effects on the phase diagram of the confined rectangular rods. Therefore the inclusion of the full rotational freedom and the in-plane positional order into the theory should be performed to prove the existence of layering transitions occurring at very high densities. However we believe that our results are qualitatively correct because Monte Carlo simulation study of freely rotating and moving hard squares confined between two parallel hard lines also shows layering transition between $n$-1 and $n$ layers [37].

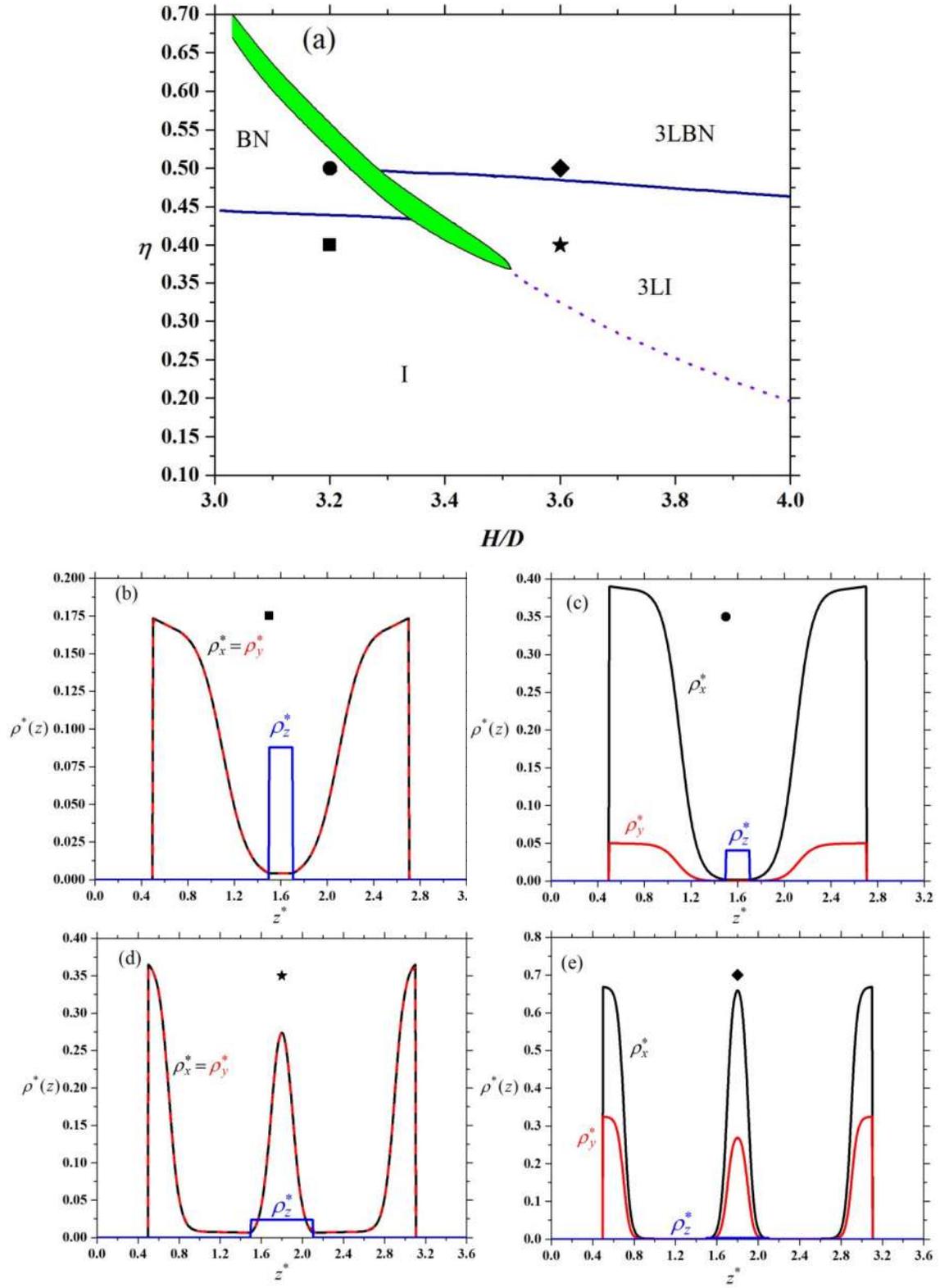

**Figure 1**

Phase diagram and observed phases of confined hard rectangular rods with *L/D*=3. The coexisting packing fractions as a function of wall-to-wall distance (a) and the density profiles of different phases from (b) to (e) are shown. The observed phases are isotropic with planar adsorption at the walls (I), surface induced biaxial nematic (BN), three-layer isotropic (3LI) and three-layer biaxial nematic (3LBN). The locations of the density profiles of (b) to (e) are marked by different symbols in the phase diagram. The dotted curve comes from the condition of $\rho_x(z=H/2)=\rho_z(z=H/2)$. The biphasic region is shaded with grey colour.

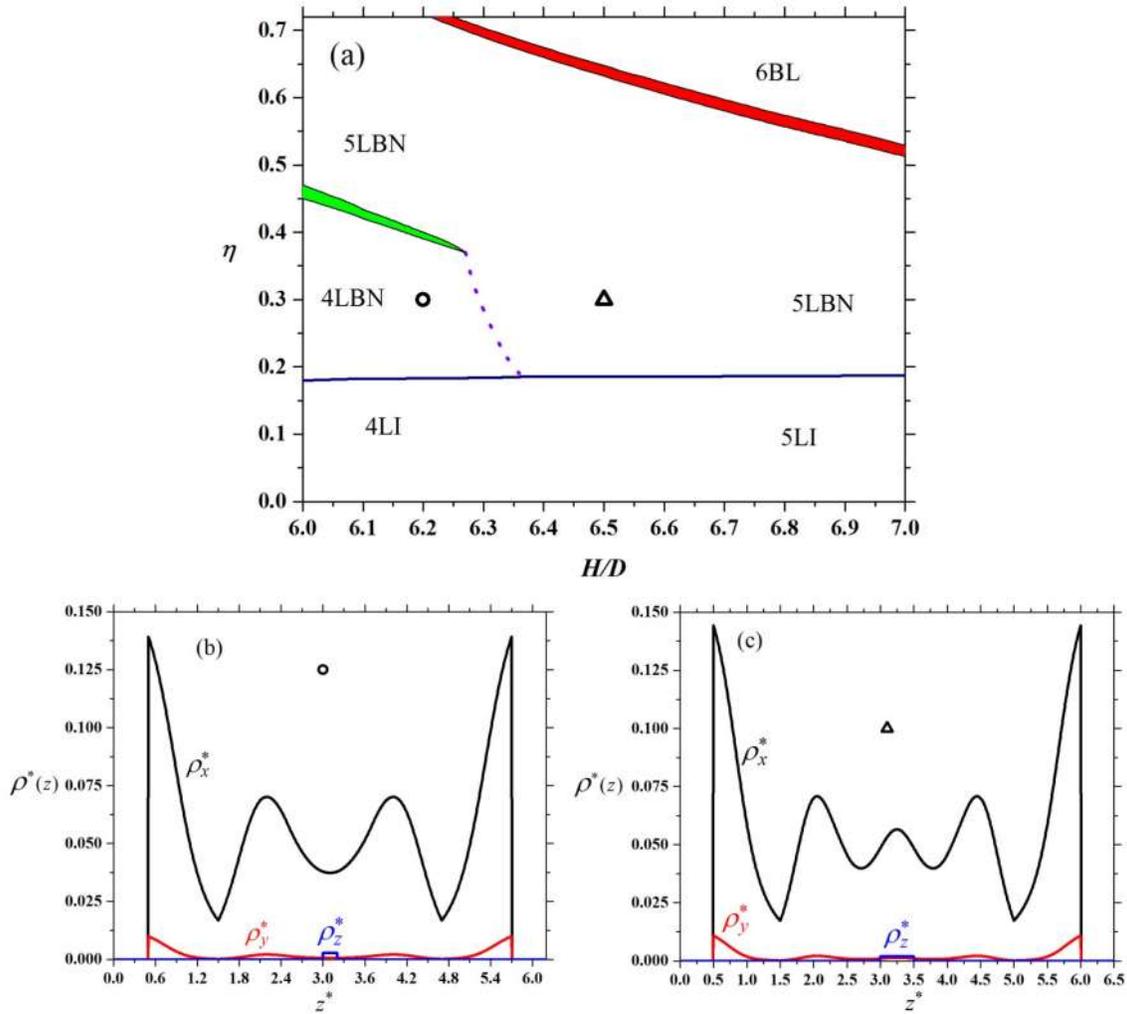

**Figure 2**

Phase diagram and observed phases of confined hard rectangular rods with *L/D*=6. The coexisting packing fractions as a function of wall-to-wall distance (a) and the density profiles of biaxial phases with four ((b) and five layers (c)) are shown. The observed phases are the isotropic with planar adsorption at the walls with four or five planar layers (4LI and 5LI) and surface induced biaxial nematic with four to six planar layers (4LBN, 5LBN and 6LBN). The locations of the density profiles of (b) and (c) are marked by different symbols in the phase diagram. The dotted curve comes from the condition of $\rho_x(z = H/2) = \rho_x(z = H/2 - \Delta z)$. The biphasic regions are shaded with different colours.

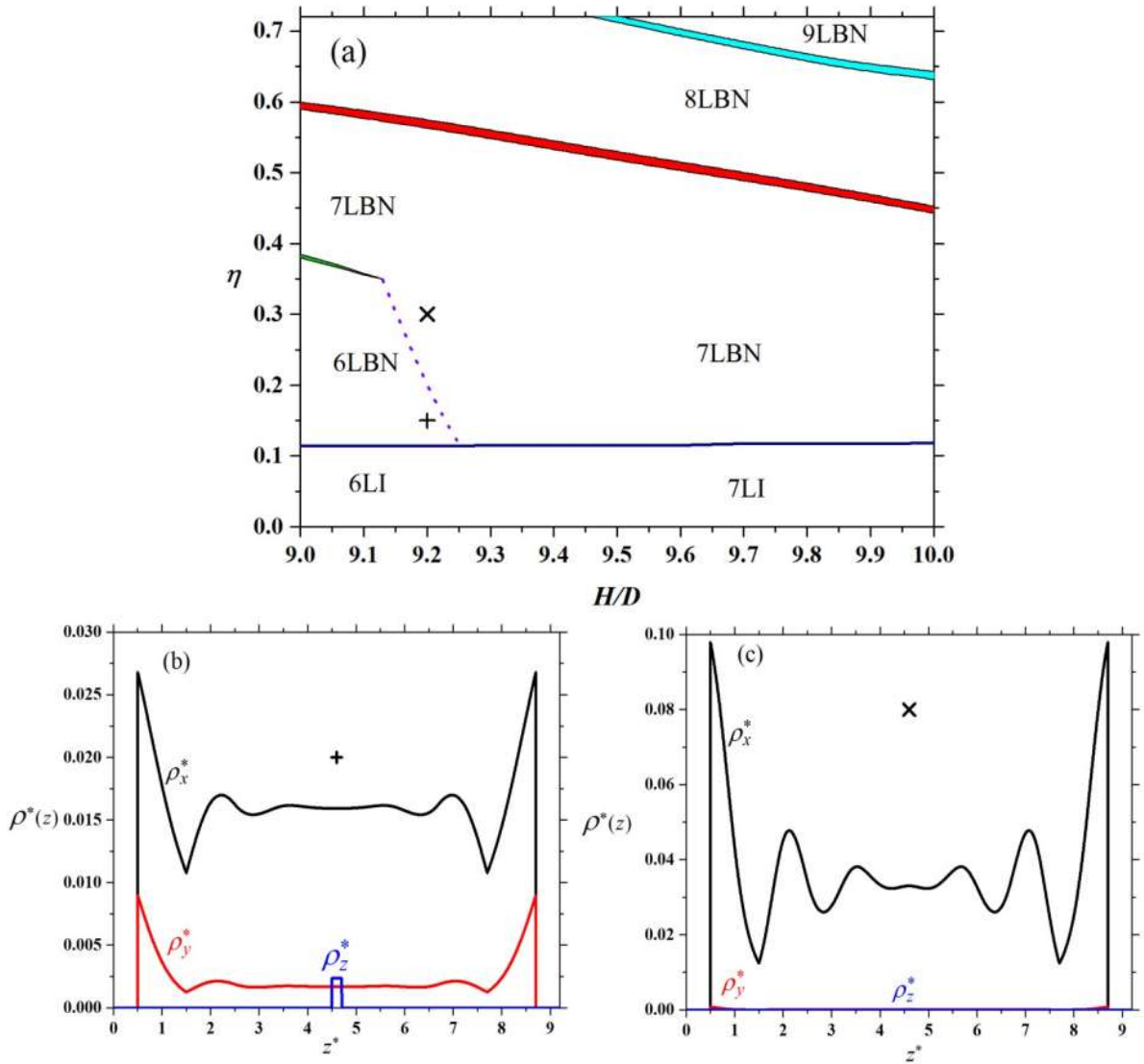

**Figure 3**

Phase diagram and observed phases of confined hard rectangular rods with $L/D=9$. The coexisting packing fractions as a function of wall-to-wall distance (a) and the density profiles of some phases ((b) and (c)) are shown. The observed phases are isotropic with planar adsorption at the walls with six or seven planar layers (6LI and 7LI) and surface induced biaxial nematic from six to nine planar layers (6LBN, 7LBN, 8LBN and 9LBN). The locations of the density profiles of (b) and (c) are marked by different symbols in the phase diagram. The dotted curve comes from the condition of $\rho_x(z=H/2)=\rho_x(z=H/2-\Delta z)$. The biphasic regions are shaded with different colours.